\documentclass[10pt,english,aps,amssymb,prb,twocolumn, showpacs]{revtex4-1}
\pdfoutput=1
\usepackage[T1]{fontenc}
\usepackage[latin9]{inputenc}
\usepackage{amsmath}
\usepackage{graphicx}
\usepackage{amssymb}
\usepackage{esint}
\usepackage{esint}
\usepackage{verbatim}

\makeatletter
\@ifundefined{textcolor}{}
{
 \definecolor{WHITE}{gray}{1}
 \definecolor{RED}{rgb}{1,0,0}
 \definecolor{GREEN}{rgb}{0,1,0}
 \definecolor{BLUE}{rgb}{0,0,1}
 \definecolor{CYAN}{cmyk}{1,0,0,0}
 \definecolor{MAGENTA}{cmyk}{0,1,0,0}
 \definecolor{YELLOW}{cmyk}{0,0,1,0}
 }

\renewcommand{\phi}{\varphi}
\renewcommand{\epsilon}{\varepsilon}

\renewcommand{\vec}[1]{{\bf #1}}

\usepackage{babel}

\begin{document}

\title {Ground state angular momentum, spectral asymmetry, and topology in chiral superfluids and superconductors}

\author{Teemu Ojanen}\email[Correspondence to ]{teemuo@boojum.hut.fi}

\affiliation{Department of Applied Physics (LTL), Aalto University, P.~O.~Box 15100,
FI-00076 AALTO, Finland }

\date{\today}
\begin{abstract}
Recently it was discovered that the ground state orbital angular momentum in two-dimensional chiral superfluids with pairing symmetry $(p_x+ip_y)^\nu$ depends on the winding number $\nu$ in a striking manner.  The ground state value for the $\nu=1$ case is $L_z=\hbar N/2$ as expected by counting the Cooper pairs, while a dramatic cancellation takes place for $\nu>1$. The origin of the cancellation is associated with the topological edge states that appear in a finite geometry and give rise to a spectral asymmetry. Here we study the reduction of orbital angular momentum for different potential profiles and pairing strengths, showing that the result $L_z=\hbar N/2$ is robust for $\nu=1$ under all studied circumstances.  We study how angular momentum depends on the gap size $\Delta/E_F$ and obtain the result $L_z=\frac{\hbar\nu}{2} N(1-\frac{\mu}{E_F})$  for $\nu=2,3$. Thus, the gap-dependence of $L_z$ for $\nu<4$ enters at most through the chemical potential while $\nu\geq4$ is qualitatively different. In addition, we generalize the spectral asymmetry arguments to  \emph{total} angular momentum in the ground state of triplet superfluids where due to a spin-orbit coupling $L_z$ is not a good quantum number. We find that the ground state total angular momentum also behaves very differently depending on total angular momentum of the Cooper pairs.

\end{abstract}
\maketitle
\bigskip{}

\section{Introduction}

In chiral superconductors and superfluids the constituent fermions are bound to Cooper pairs with non-zero angular momentum that form a macroscopic condensate. The ground state orbital angular momentum in chiral superfluids, such as $^3$He-A, has been under debate for over 40 years.\cite{volovik1,vollhart, leggett2, anderson,  leggett1, volovik4, cross, ishikawa, mermin, volovik5, balatsky,stone1,sauls,tada,volovik2}  Over the years different arguments and methods have resulted in dramatically different predictions. In a nutshell, there seem to be a contradiction between two simple physical pictures. On one hand the ground state of $N$ fermions can be thought of as a collection of $N/2$ Cooper pairs with angular momentum $\nu \hbar$ where $\nu$ is an integer that describes the orbital state of a Cooper pair. This leads to the estimate  $L_z=\nu\hbar N/2$ for the orbital angular momentum in the ground state. On the other hand, the superconducting ground state in the BCS theory is described by a modification of the free fermion ground state in an energy shell $\Delta$, outside which the system is roughly unchanged. This point of view seems to imply that only the states in the pairing energy window $\Delta$ contribute to the angular momentum $L_z\sim N(\frac{\Delta}{E_F})^\gamma$, where $\gamma>0$ and $E_F$ is the Fermi energy. Since the pairing energy may be much smaller than the Fermi energy $\Delta\ll E_F$, the second estimate for the ground state angular momentum is dramatically reduced. A reconciliation between these contradictory conclusions have remained elusive.

\begin{figure}
\includegraphics[width=0.9\columnwidth]{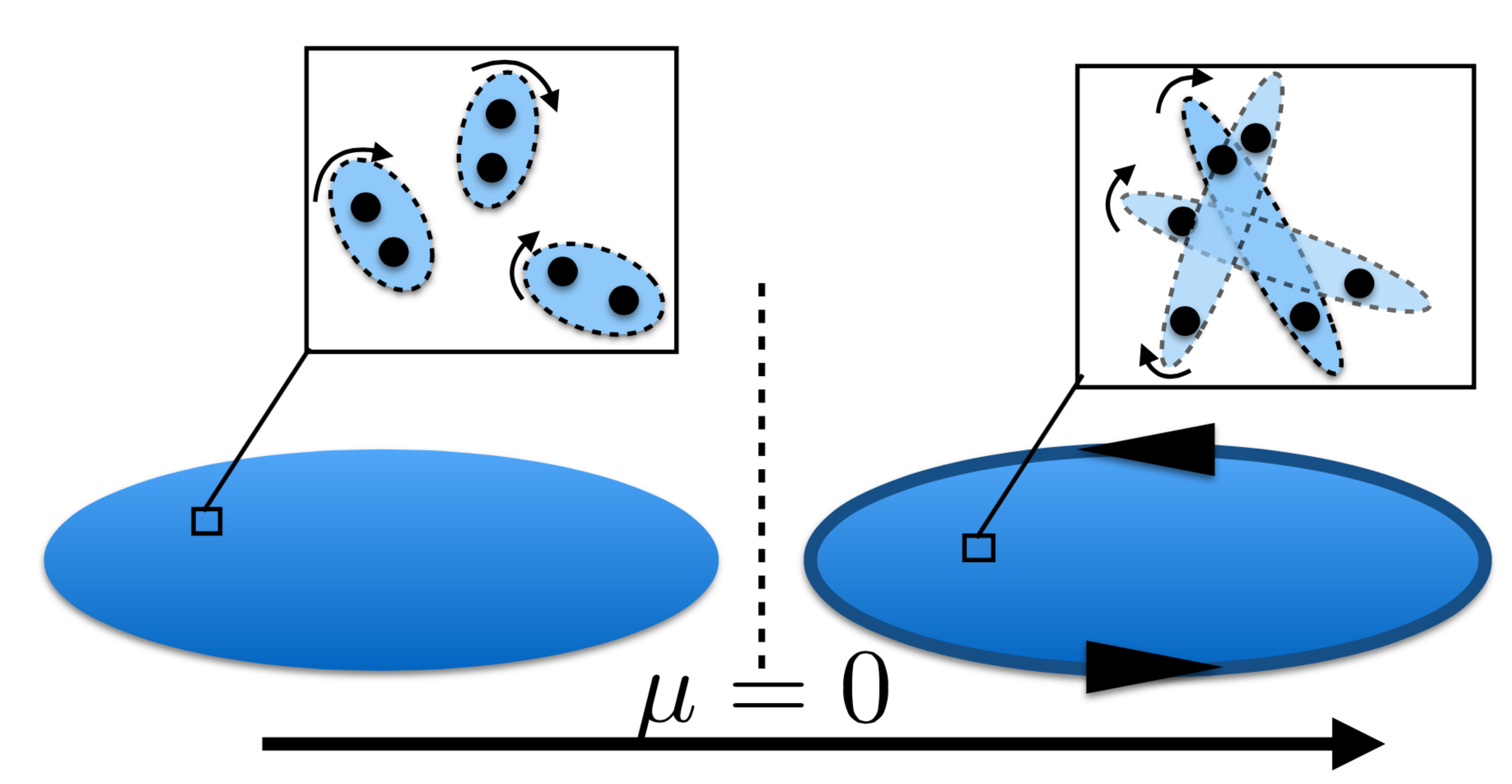}
\caption{Chiral fermionic superfluids with $(p_x+ip_y)^\nu$ pairing symmetry undergo a topological phase transition when the chemical potential crosses zero. On the negative side, the system is in a trivial strong pairing phase while a positive chemical potential corresponds to a topological weak pairing superfluid. The phase transition is accompanied by a closing of the energy gap and an appearance of chiral edge modes.  The edge modes have opposite chirality compared to the Cooper pairs and and strongly modify the ground state angular momentum for $\nu\geq2$. }
\label{scematic}
\end{figure}
Recent insights based on topology and edge states in a finite geometry have lead to important new understanding in the angular momentum problem. In the standard BCS description, the number of particles $N$ and analogous one-body observables are not conserved quantities due to the $U(1)$ symmetry breaking pairing terms. Building on Volovik's notion that the operator combination $\hat{L}_z-\nu \hat{N}/2$ still commutes with a rotationally invariant Hamiltonian with $(p_x+ip_y)^\nu$ pairing terms, Tada, Nie and Oshikawa expressed the ground state angular momentum in a disc geometry in terms of a spectral asymmetry index of the Bogoliubov- de Gennes Hamiltonian.\cite{tada} A finite spectral asymmetry index, studied by Volovik\cite{volovik1, volovik3} in the context of $^3$He, arises from a spectral flow taking place in the topological phase transition between a strong-pairing phase and a weak-pairing phase when the chemical potential crosses zero,\cite{read1} see Fig.~\ref{scematic} The spectral flow is generated by the edge states present in the weak pairing phase, thus being a consequence of nontrivial momentum space topology and finite geometry. 

The new approach by Tada \emph{et al.}\cite{tada} has confirmed that the ground state angular momentum of s $p_x+ip_y$ superfluid is indeed  $L_z=\hbar N/2$ as expected by counting the Cooper pairs. Even though the system supports a chiral edge state, the spectral flow in the topological phase transition stays zero and the asymmetry index vanishes. Remarkably, superfluids with chiral pairing $(p_x+ip_y)^\nu$, $\nu\geq2$ the situation is drastically different. In that case the edge currents lead to dramatic cancellation of the bulk contribution and $L_z=0$ when $\Delta/E_F\to 0$. This result holds even in the thermodynamic limit, in contrast to conventional thinking which would  suggest that the boundary effects become insignificant.

In the current work we generalize the work of Ref.~\onlinecite{tada} in various aspects.  We study the spectral asymmetry index and the ground state angular momentum on a disc with a hard-wall confining potential as a function of the pairing gap $\Delta$ for different $(p_x+ip_y)^\nu$ pairing states. We conclude that the universal behaviour of the spectral asymmetry index discovered in Ref.~\onlinecite{tada} in the limit $\Delta/E_F\to 0^+$ extends also to finite $\Delta/E_F$ ratio when $\nu<4$. We obtain that, within the accuracy of our numerics, the correction to the naive estimate $L_z=\nu\hbar N/2$, which always holds for $\nu=1$, is universal and leads to relation $L_z=\frac{\nu}{2} N(1-\frac{\mu}{E_F})$. In the small-gap limit it holds that $\mu\to E_F$ and $L_z=0$.  Our findings imply that for $\nu=2,3$ the angular momentum suppression depend on the gap \emph{only} through $\mu$ and not extra suppression factors $(\frac{\Delta}{E_F})^\gamma$ should arise. We also compare the effects of smooth and sharp confining potentials on the ground state angular momentum. This is particularly interesting since it has been proposed that the potential profile near the edge could lead to different estimates for $L_z$ and explain parts of the angular momentum controversy.\cite{stone1, huang, huang2, goryo}  In qualitative agreement with previous works, we find that the ground state angular momentum for $\nu=1$ case is independent of the shape of the confining potential and is given by the universal value $L_z=\hbar N/2$, whereas for $\nu>1$ the ground state value of $L_z/N$ can be widely tuned by the confining potential.

In the second part of the work we will consider total angular momentum in chiral superconductors where the orbital angular momentum is not a good quantum number.  Analogously to the operator $\hat{L}_z-\nu \hat{N}/2$ in the absence of a spin-orbit coupling, we can prove that for certain odd-pairing superfluids the quantity $\hat{J}_z-j_z\hat{N}/2$ commutes with the Hamiltonian. Here $j_z$ denotes the total angular momentum of a Cooper pair. Therefore the $U_{J_z}\times U(1)_{N}$ rotational symmetry in real and wavefunction space  reduces to a combined symmetry $U_{J_z-j_zN/2}$ in the presence of pairing correlations. This notion allows us to express total angular momentum in terms of the spectral asymmetry index in the ground state and generalize the results of Ref.~\onlinecite{tada} to the case where  orbital angular momentum is not a good quantum number even in the absence of superconductivity. We will demonstrate that the spectral asymmetry index behaves very differently depending on the total angular momentum of a Cooper pair. Below we will adopt the units where $\hbar=1$, although $\hbar$ is restored in some statements for clarity.

\section{Chiral superfluids on  a disc}

In this work we will study two-dimensional chiral superconductors and superfluids described by the second-quantized Hamiltonian 
\begin{align} \label{h1}
&\hat{H}=\hat{h}+\hat{\Delta},
\end{align}
where
\begin{align} \label{h2}
&\hat{h}=\int d^2r \hat{\psi}_{\alpha}^\dagger(\vec{r})\left[\epsilon_k\delta_{\alpha\beta}+\vec{b}_{\alpha \beta}(\vec{k})\right]\hat{\psi}_\beta(\vec{r}),\nonumber\\
&\hat{\Delta}=\int d^2r \left[\hat{\psi}_{\alpha}(\vec{r})\Delta_{\alpha\beta}(\vec{k})\hat{\psi}_\beta(\vec{r})+\mathrm{h.c}\right].
\end{align}
The field operators satisfy fermionic anticommutation relations $\{ \hat{\psi}_{\alpha}(\vec{r}) ,\hat{\psi}_{\beta}(\vec{r}') \}=0$, $\{ \hat{\psi}_{\alpha}(\vec{r}) ,\hat{\psi}_{\beta}^{\dagger}(\vec{r}') \}=\delta_{\alpha\beta}\delta(\vec{r}-\vec{r}')$. The spin-diagonal part $\epsilon_k=\left[\frac{k_x^2+k_y^2}{2m}-\mu+V(r)\right]$ includes kinetic energy, chemical potential and a confining potential.  The spin-dependent energy $\vec{b}=\alpha_R(k_x\sigma_y-k_y\sigma_x)+B_z\sigma_z$ is determined by a Rashba spin-orbit coupling and a perpendicular Zeeman field and the pairing correlations are encoded in the gap function $\Delta_{\alpha\beta}(\vec{k})$. In this work we will focus on chiral pairings with orbital structure $(k_x\pm ik_y)^\nu$ and employ substitution $\vec{k}=-i\nabla$. The Pauli matrices $\sigma_i$ and the subscripts $\alpha, \beta$ and  are associated with the spin degrees of freedom and repeated indices imply summation. In the following we will assume that the potential $V(r)$ depends only on the radial coordinate that confines the particles to a disc with radius $R$. Further, we assume that the potential vanishes for $r<R$, with the exception of Sec.~\ref{potential} where we will consider smooth confining potentials. The second-quantized operators for the orbital angular momentum $\hat{L}_z$, the perpendicular spin $\hat{S}_z$, and the particle number $\hat{N}$  are given by 
\begin{align} \label{oper}
&\hat{L}_z=\int d^2r \hat{\psi}_{\alpha}^\dagger(r)\left[xk_y-yk_x\right]\hat{\psi}_\alpha(r) \nonumber \\
&\hat{S}_z=\int d^2r \hat{\psi}_{\alpha}^\dagger(r)\frac{1}{2}\left[\sigma_z\right]_{\alpha\beta}\hat{\psi}_\beta(r)  \nonumber \\
&\hat{N}=\int d^2r \hat{\psi}_{\alpha}^\dagger(r)\hat{\psi}_\alpha(r).
\end{align}
In addition, the total angular momentum operator is given by $\hat{J}_z=\hat{L}_z+\hat{S}_z$.

We will study the ground state properties of the Hamiltonian in Eq.~(\ref{h1}).  To this end we will define the associated $4\times4$ Bogoliubov-de Gennes (BdG) Hamiltonian 
\begin{align} \label{bdg1}
H(k)=\begin{pmatrix} h(\vec{k}) & \Delta(\vec{k}) \\ -\Delta^* (-\vec{k})& -h^T(\vec{-k}) \end{pmatrix} ,
\end{align}
where $h(k) =\epsilon_k+\vec{b}_{\alpha \beta}(\vec{k})$ and the off-diagonal block is given by the pairing matrix $\Delta_{\alpha\beta}(\vec{k})$. The wavefunctions are four-component spinors $\psi_(\vec{r})=\left[u_{\uparrow}(\vec{r}), u_{\downarrow}(\vec{r}),v_{\uparrow}(\vec{r}), v_{\downarrow}(\vec{r})\right]^T$. We will solve the eigenvalue problem in the position representation and work in the polar coordinates  $(r,\phi)$ where $\nabla^2=\partial_r^2+\frac{1}{r}\partial_r+\frac{1}{r^2}\partial_\phi^2$ and $k_x\pm ik_y=-i e^{\pm i \phi}(\partial_r\pm \frac{i}{r}\partial_\phi)$. Since the BdG Hamiltonian (\ref{bdg1}) does not contain potential terms that would  explicitly depend on the polar angle, we seek to factor out the angular dependence in the eigenfunctions.  To carry out the separation of variables, we will pass to a new basis $\psi'(r)=U(\phi)\psi(r,\phi)$ where the spinors only depend on the radial coordinate. The  unitary transformation is a diagonal $4\times4$ matrix of the form $U^{\dagger}(\phi)=e^{il\phi}\mathrm{diag}[e^{iN_1\phi}, e^{iN_2\phi},e^{iN_3\phi},e^{iN_4\phi}]$, where the integers $N_i$ can be determined when the specific pairing symmetry is chosen.  We will give  the explicit form of $U(\phi)$  for the different cases below. Since the transformation depends on the angular quantum number $l$, the BdG Hamiltonian becomes $H'=UHU^{\dagger}\equiv H(r,l)$ and the eigenvalue problem takes the form
\begin{align} \label{bdg2}
H(r,l)\psi_{nl}(r)=E_{n,l}\psi_{nl}(r).
\end{align}
where the eigenfunctions $\psi_{nl}(r)=\left[u_{nl\uparrow}(r), u_{nl\downarrow}(r),v_{nl\uparrow}(r), v_{nl\downarrow}(r)\right]^T$ are labelled by the radial and angular quantum numbers  $n$ and $l$. To study the properties of a many-body ground state we need to solve the $r$-dependent one-dimensional eigenvalue problem for all different $l$. The eigenvalue problem is fully determined with the boundary condition $\psi_{nl}(R)=0$ together with the smoothness condition for derivatives at $r=0$. We impose the standard normalisation condition
\begin{align}
2\pi\sum_\sigma \int_0^R dr r\left[  |u_{nl\sigma}(r)|^2+|v_{ln\sigma}(r)|^2 \right]=1 .\nonumber
\end{align}
The ground state expectation value of the particle number, orbital angular momentum and spin operators in the ground state can be expressed as 
\begin{align} \label{lz}
&L_z=-2\pi  \times \nonumber  \\ 
& \sum_{E_{nl}>0} \int_0^R  dr r \left[ (l+N_3) |v_{nl\uparrow}(r)|^2+(l+N_4)|v_{ln\downarrow}(r)|^2 \right], \nonumber \\
&S_z=2\pi \sum_{E_{nl}>0}\int_0^R  dr r \frac{1}{2}\left[ |v_{nl\uparrow}(r)|^2-|v_{ln\downarrow}(r)|^2 \right], \nonumber \\
&N=2\pi \sum_{E_{nl}>0}\int_0^R  dr r \left[  |v_{nl\uparrow}(r)|^2+|v_{ln\downarrow}(r)|^2 \right].
\end{align}
The summation is performed over all positive energy states labelled by the quantum numbers $n$ and $l$.
For high-order pairing $\Delta(\vec{k})\propto(k_x+ik_y)^\nu$ with $\nu\geq 2$, one needs to introduce an explicit high-energy cutoff  above which $\Delta(\vec{k})=0$ to render $N$, $L_z$ and $S_z$ finite. Physically, this means that the Cooper pairing is taking place within a finite energy window.

\section{Orbital angular momentum and spectral asymmetry in chiral superfluids}

\subsection{Angular momentum and spectral asymmetry index} \label{oam}
In this section we will consider simple chiral superfluids where the spin serves mostly as an inert spectator.  We assume that the Zeeman and the Rashba coupling vanish, set $\vec{b}_{\alpha\beta}=0$ in Eqs.~(\ref{h1}), (\ref{bdg1}), and concentrate on pairing symmetries of the form
\begin{align} \label{delta1}
\Delta_{\alpha\beta}(\vec{k})=\Delta_0\left( k_x+ik_y \right)^\nu\delta_{\alpha\uparrow}\delta_{\beta\downarrow},
\end{align}
where $\nu$ is an integer.\cite{tada} The even $\nu$ case corresponds to spin singlet pairing and the odd $\nu$ case corresponds to a triplet pairing. The usual triplet parametrisation  $\Delta(\vec{k})=i\sigma_\mu d_\mu(\vec{k})\sigma_y$ implies that the $d$ vector is of the form $d_\mu=(0, 0, d_z)$.  The $\nu=1$ case  could represents a thin film of $^3$He-A or Sr$_2$RuO$_4$.\cite{volovik1, mackenzie} These systems do not exhaust the list of chiral superfluids -- in addition to the intrinsic chiral superconductors and superfluids, there exists various schemes to engineer chiral superconductors by combining simple systems into heterostructures.\cite{qi} For example, recently it was discovered that superconductors with high Chern numbers could be realized in planar systems with magnetic adatoms.\cite{ront1, ront2} Also, a recent proposal suggests that Sr$_2$RuO$_4$ would actually be described by a Chern number 7 state. 

A continuous symmetry generally implies a conserved quantity which serves as a generator of the symmetry transformation. Angular momentum is the generator of rotations and rotational symmetry about the $z$ axis typically implies that the system Hamiltonian commutes with the angular momentum operator $\hat{L}_z$ (it is sufficient to consider the orbital part of $\hat{J}_z$ when the spin does not play a role). However, the Hamiltonian (\ref{h1}) with the gap function (\ref{delta1}) does not commute with $\hat{L}_z$ due to the pairing terms. This failure of commuting has a similar origin to the BCS-type $U(1)$ symmetry breaking resulting in the well-known particle nonconservation. As observed by Volovik,\cite{volovik1,volovik3} it is possible to find a conserved quantity by defining the operator  
\begin{align} \label{what}
\hat{Q}\equiv \hat{L}_z-\frac{\nu}{2}\hat{N}
\end{align}
which indeed satisfies $[\hat{Q},\hat{H}]=0$.  The original symmetry $U(1)_{L_z}\times U(1)_{N}$ of a rotationally invariant and particle conserving normal system reduces to a single combined symmetry $U(1)_{L_z-\nu N/2}$ when the system becomes superfluid and the anomalous terms in Eq.~(\ref{h1}) become finite.

The ground state of the superfluid can be constructed by filling the negative energy Bogoliubov quasiparticle states  obtained as solutions to the BdG eigenvalue problem (\ref{bdg2}). Due to particle-hole symmetry, this is equivalent to saying that all the positive energy quasiparticle states are empty.  In this case, the unitary transformation $U(\phi)$ leading to the separable eigenvalue problem (\ref{bdg2}) is fixed by the integers $N_1=N_2=\nu$ and $N_3=N_4=0$.   Each negative-energy state $\psi_{nl}(r)$ gives a contribution $(l+\nu/2)/2$ to the ground state eigenvalue of $\hat{Q}$. Summing up all the occupied states, the ground state value is given by
\begin{align} \label{Q}
Q=-\frac{1}{2}\sum_{l} (l+\frac{\nu}{2})\eta_l,
\end{align}
where $\eta_l=\frac{1}{2}\sum_n \text{sign}\,E_{n,l}$, which is called the spectral asymmetry index, measures the difference between the number of negative and positive energy solutions of the BdG eigenvalue problem (\ref{bdg2}) for a fixed $l$.\cite{tada} The appearance of the spectral asymmetry in Eq.~(\ref{Q}) follows from the particle-hole symmetry of the spectrum $E_{n,l}=-E_{n,-l-\nu}$ which implies that the number of negative energy states at $l$ is equal to the positive energy states at $-l-\nu$. The particle-hole symmetry of the eigenvalues follows from the particle-hole symmetry of the BdG Hamiltonian $P H(r,l)P^{-1}=-H^*(r,-l-\nu)$, where $P=\tau_x$ ($P=i\tau_y$) for even (odd) $\nu$. 

\begin{figure}
\includegraphics[width=0.99\columnwidth]{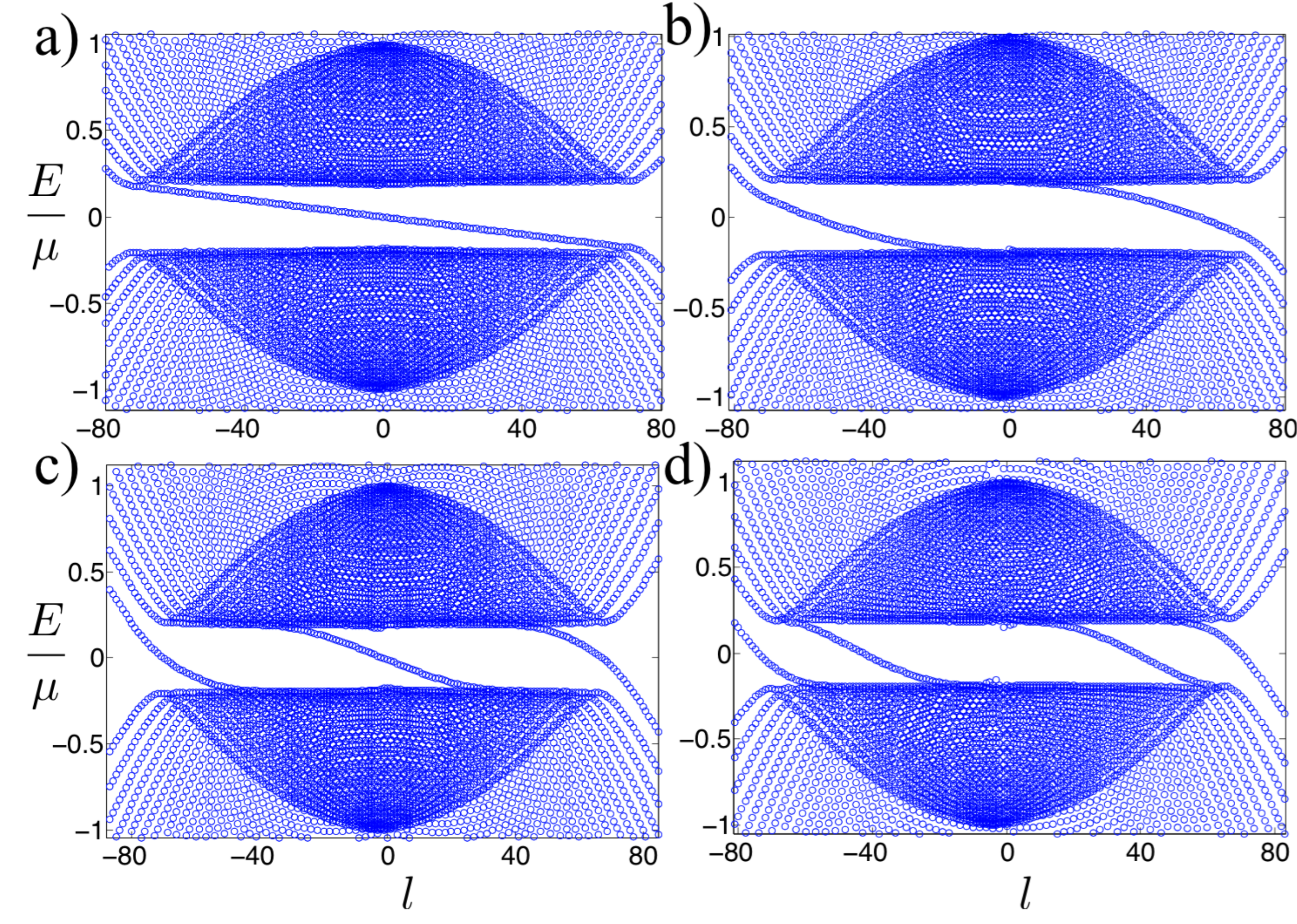}
\caption{Spectra of chiral superfluids with  different pairing symmetries $\nu$ as a function of the angular quantum number $l$. Figures  a)-d)  represent cases $\nu=1-4$. The other parameters are $k_F^\nu\Delta_0=0.2E_F$, $\mu=0.9E_F$ and $R=80k_F^{-1}$. The states traversing the bulk gap are localized to the edge of the disc.}
\label{wind}
\end{figure}

In Fig.~\ref{scematic} we have schematically illustrated how chiral superfluids with the pairing term (\ref{delta1}) exhibit a topological phase transition at $\mu=0$ from a strong-pairing phase of Bose-condensed molecules $\mu<0$ to a weak-pairing BCS state $\mu>0$ of loosely bound pairs.  As was pointed out in Refs.~\onlinecite{tada, volovik2}, the spectral asymmetry is generated by the spectral flow in the topological phase transition. The spectral asymmetry is intimately related to the chiral edge states appearing in the BCS regime  in the presence of a boundary. The ground state expectation values satisfy equation $L_z=\frac{\nu}{2}N+Q$, therefore $Q$ can be interpreted as the reduction of the orbital angular momentum from the value $\frac{\nu \hbar}{2}$ per Cooper pair in the $\mu<0$ regime  due to the edge effects. The existence of the edge states is guaranteed by topology since for $\mu>0$ and the pairing term (\ref{delta1}), the Chern number is given by $2\nu$, which is in agreement with the existence of $\nu$ doubly degenerate chiral edge mode branches in the BCS regime. In Fig.~\ref{wind} we have plotted the spectrum as a function of the angular quantum number $l$, that clearly illustrates the existence of the edge modes.  As discussed in Refs.~\onlinecite{tada} and \onlinecite{volovik2},  the angular momentum reduction $Q$ is determined by the values $l_\nu$ for which the edge dispersion crosses the zero energy.  This is seen in Fig.~\ref{asy}, illustrating that the changes in the spectral asymmetry may take place only at the crossings of the zero energy and $l=-\nu/2$ due to particle-hole symmetry.

\subsection{Ground state angular momentum and the pairing gap}

In Ref.~\onlinecite{tada} it was concluded that the ground state orbital angular momentum reduction in the $\mu>0$ regime for $\nu=1$ is always absent and $L_z=\frac{1}{2}N$, while for $\nu\geq 2$ and $k_F^\nu\Delta_0\ll E_F$ it is $Q= -\nu Q_0$, where  $Q_0=\frac{1}{4}R^2k_F^2$. Here the Fermi momentum is given by the expression $k_F=\sqrt{2mE_F}$ and the Fermi energy is determined by the total number of particles at vanishing gap. This result indicates that the angular momentum suppression for $\nu\geq 2$ is essential and the boundary effects modify the bulk result  dramatically even in the thermodynamic limit. Since the particle number is given by the expression $N=\frac{R^2k_F^2}{2}$, Ref.~\onlinecite{tada} arrived at the remarkable conclusion that $L_z=0$  for $\nu \geq2$ and small gap  $k_F^\nu\Delta_0\ll E_F$.   
\begin{figure}
\includegraphics[width=0.99\columnwidth]{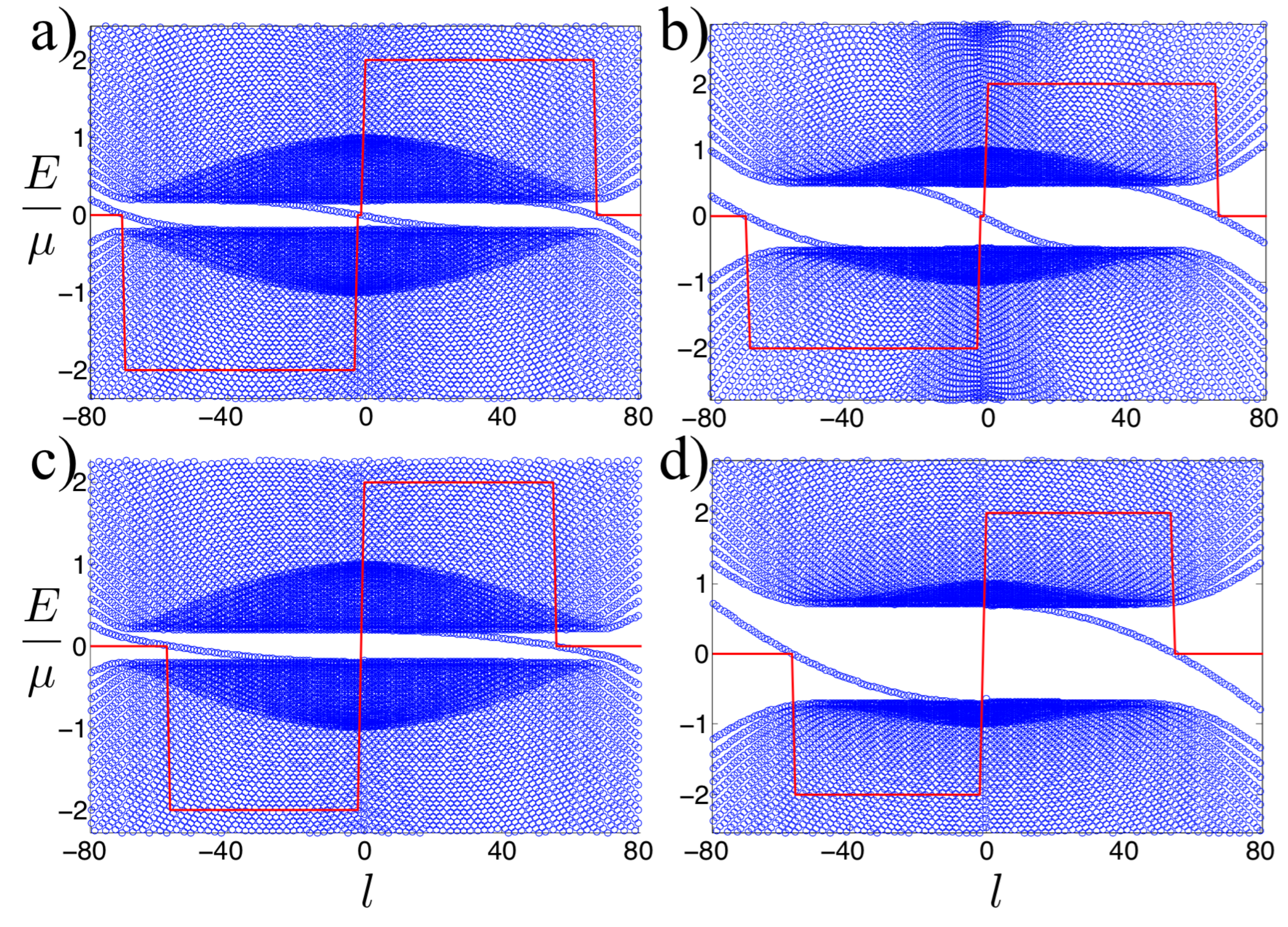}
\caption{a): Spectrum of $\nu=3$ superfluid  as a function of the angular quantum number $l$ with $k_F^3\Delta_0=0.2E_F$, $\mu=0.9E_F$ and $R=80k_F^{-1}$. b): The same as a) but for larger gap  $k_F^3\Delta_0=0.9E_F$. c): The same as  a) but for pairing $\nu=2$ and $k_F^2\Delta_0=0.2E_F$, $\mu=0.9E_F$. d): The same as c) but for a larger gap $k_F^2\Delta_0=0.9E_F$, $\mu=0.8E_F$. The red line denotes the spectral asymmetry $\eta_l$. The numerical value of $\eta_l$ coincides with the $E/\mu$ axis value.}
\label{asy}
\end{figure}

\begin{figure}
\includegraphics[width=0.8\columnwidth]{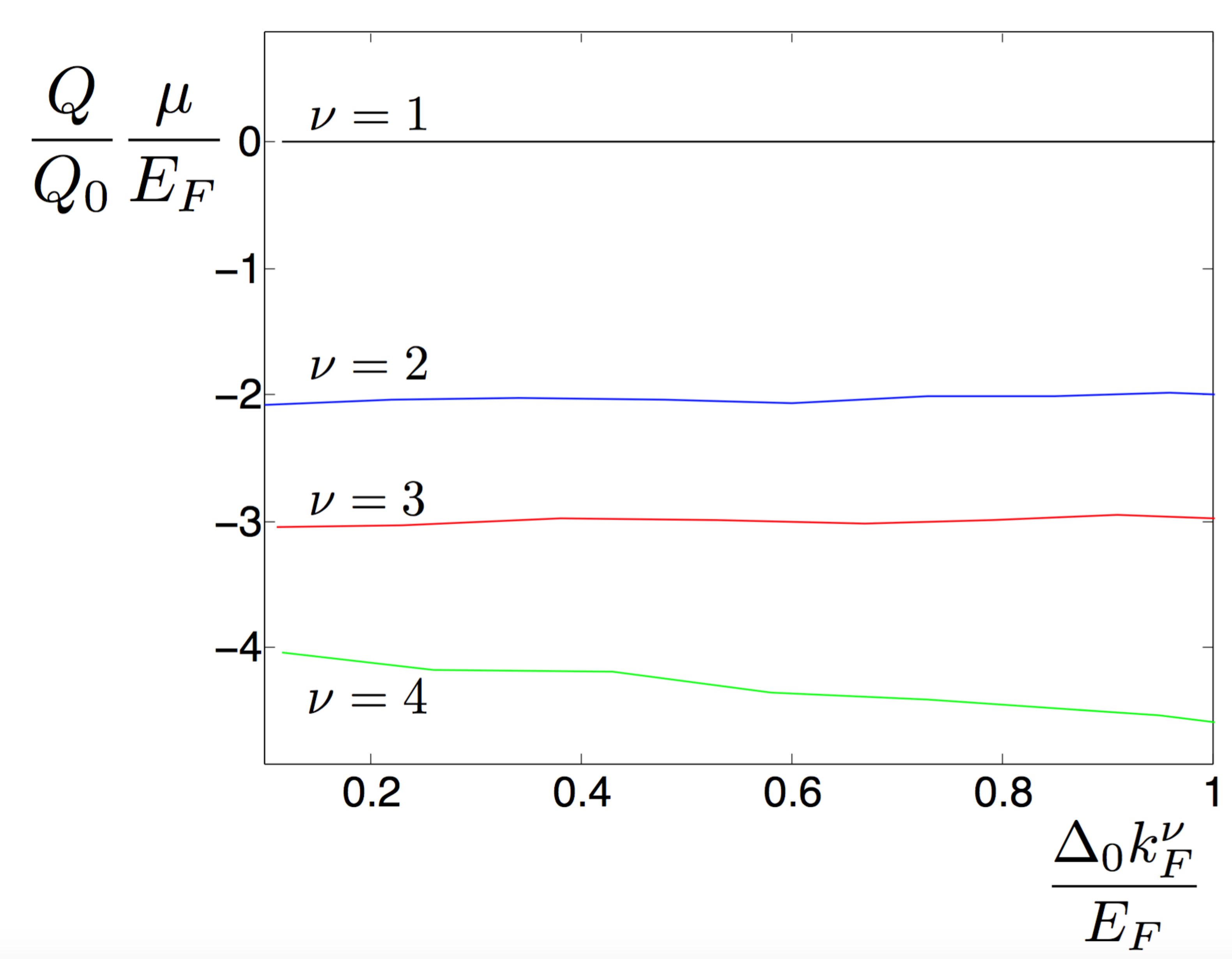}
\caption{Angular momentum reduction factor for different pairing symmetries as a function of the paring gap for $R=80k_F^{-1}$ and fixed $N$. We see that $Q\frac{\mu}{E_F}$ is independent of the pairing strength for $\nu=1, 2,3$, while $\nu=4$ is qualitatively different. }
\label{Q1}
\end{figure}

Now we will explore how the reduction factor $Q$ behaves outside the small gap regime. To this end, we diagonalize $\hat{H}$ on a disc and evaluate expression (\ref{Q}). We study weak pairing phase $\mu>0$ and keep $N$ fixed while increasing $k_F^\nu \Delta_0/E_F$. Chemical potential is then determined by  $N$ and $k_F^\nu \Delta_0/E_F$.    In Fig.~\ref{Q1} we have plotted $Q$ as a function of the pairing strength.  As discussed above, in the small-gap limit $Q$ is determined by a single parameter $Q_0$. Remarkably, our calculations imply that for $\nu=2,3$ and $k_F^\nu \Delta_0\sim E_F$ similar results holds for a scaled quantity quantity $Q\frac{\mu}{E_F}$. Therefore for $\nu=2,3$ we obtain that $L_z=\frac{\nu}{2} N(1-\frac{\mu}{E_F})$  valid even for $k_F^\nu \Delta_0/E_F\sim 1$. For small pairing amplitude $\mu\approx E_F$ and the $L_z=0$ as discussed in Ref.~\onlinecite{tada}.  The reason for the universal behaviour of $L_z$ is illustrated in Fig.~\ref{asy}, the zero energy crossing of the edge branches for $\nu=2,3$ remain fixed even when $k_F^\nu\Delta_0/\mu$ is increased to large values. Even though the gap changes substantially, the spectral asymmetry remains practically unchanged in the process. Therefore, the only dependence of $Q$ on the gap amplitude must arise because $\mu$ changes when $k_F^\nu \Delta_0/E_F$ is increased with $N$ kept fixed. Generally $\mu$ as a function of $N$ and $k_F^\nu \Delta_0$ will depend on $\nu$ and how the pairing term is cut off at large energies, but relation $L_z=\frac{\nu}{2} N(1-\frac{\mu}{E_F})$ is independent of these details. 

\begin{figure}
\includegraphics[width=0.99\columnwidth]{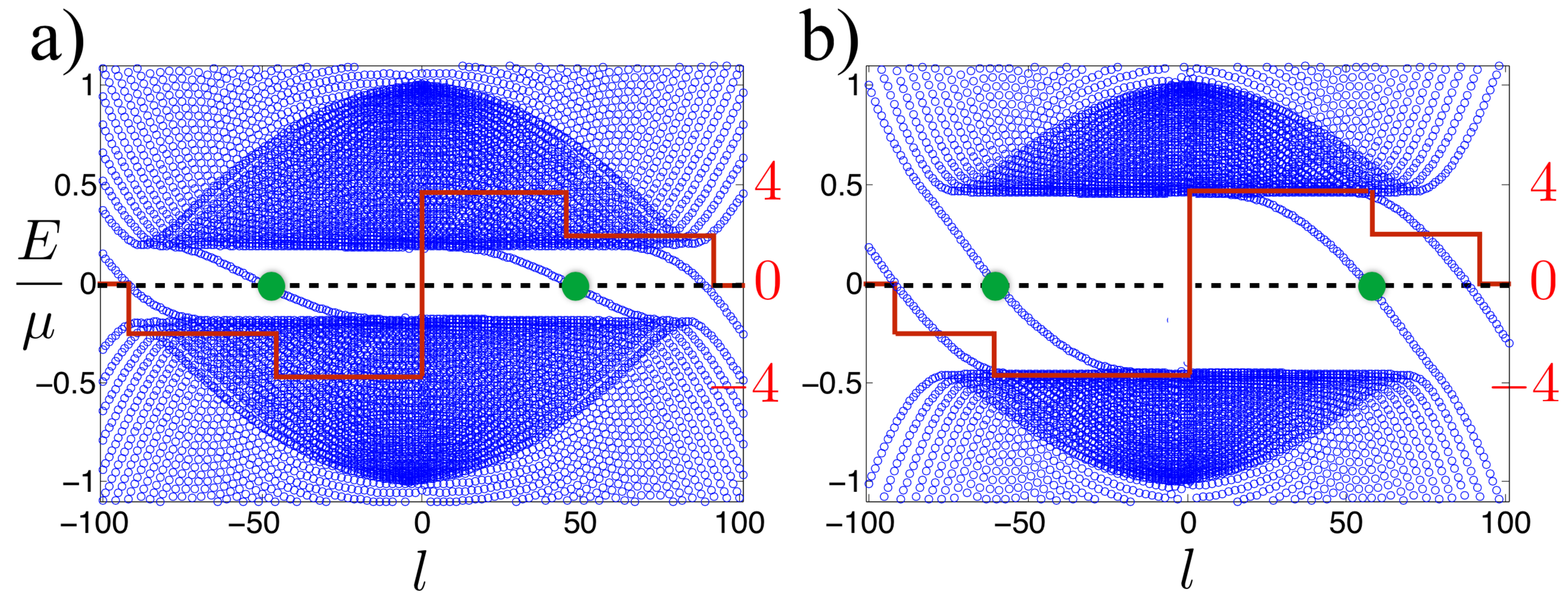}
\caption{a) Spectrum of the $\nu=4$ state as a function of the angular quantum number $l$ for  $k_F^4\Delta_0=0.3E_F$ and $R=80k_F^{-1}$. b) same as a)  but for larger gap $k_F^4\Delta_0=0.9E_F$. The red line denotes the spectral asymmetry $\eta_l$. The spectral asymmetry changes because the zero-energy intersection of the inner wings of edge dispersion (represented by the green dots), move away from $l=0$. }
\label{nu4}
\end{figure}

The $\nu=4$ case behaves qualitatively differently from $\nu=2,3$.  As illustrated in Fig.~\ref{Q1}, the result $Q\approx -\nu Q_0$ and thus $L_z=0$ holds in the small-gap limit.  When the pairing amplitude is increased, the angular momentum reduction becomes stronger $|Q\frac{\mu}{E_F}|>\nu Q_0$. This can be understood by inspecting the spectra in Fig.~\ref{nu4}, which shows that the spectral asymmetry changes because the zero energy crossing of the inner edge branches move when the gap is increasing. The outer edge branches, similar to those seen in the $\nu=2$ and $\nu=3$ cases, are essentially fixed. However, the zero energy crossings of the inner branches move symmetrically away from $l=0$ so that the spectral asymmetry and $|Q\frac{\mu}{E_F}|$ increase. Since the higher pairing states $\nu>4$ exhibit qualitatively similar inner edge branches compared to those that are responsible for the increase of the spectral asymmetry, we assume that $|Q\frac{\mu}{E_F}|$ is  an increasing function of the pairing amplitude for all  $\nu>4$. If $|Q|>\nu Q_0$ it is possible  that the ground state angular momentum changes sign as a function of $\frac{k_F^\nu \Delta_0}{E_F}$. Whether this happens or not depends also on the behaviour of $\frac{\mu}{E_F}$ which is not universal in the sense that besides $\nu$, $N$ and $k_F^\nu \Delta_0/E_F$, it also depend on the high energy cutoff of the pairing term. Since the number of the inner edge branches that give rise to increasing spectral asymmetry, illustrated for $\nu=4$ in Fig.~\ref{nu4}, increase by two when $\nu$ increase by three, the sign reversal should be more probable for increasing $\nu$. The sign reversal of the ground state angular momentum has been previously discovered in different contexts.\cite{tsutsumi2, bouhoun} 
   
We have also checked the relation $L_z-\frac{\nu}{2}N=Q$ by evaluating the left-hand side directly by employing Eq.~(\ref{lz}) without resorting to Eq.~(\ref{Q}). In this approach one needs to impose an energy cutoff for the pairing term when $\nu\geq 2$ to render $L_z$ and $N$ finite. Therefore $L_z$ and $N$  are cutoff-dependent quantities. Nevertheless, the cutoff dependence disappears from the combination  $L_z-\frac{\nu}{2}N$ which matches $Q$ calculated from Eq.~(\ref{Q}) that depends only on the spectral asymmetry, changes of  which is determined by the spectrum in the vicinity of zero energy. 

Finally, the results of this section should be contrasted with the predictions stating that orbital angular momentum in chiral superfluids is suppressed as $L_z\sim N(\Delta/E_F)^\gamma$, with $\gamma=1$\cite{anderson, leggett2} or  $\gamma=2$.\cite{volovik4,cross}  Our results imply that the combination $L_z-\frac{\nu}{2}N$ is independent of $(\Delta/E_F)$ for $\nu=1$. In addition, for $\nu=2,3$ it follows that $L_z-\frac{\nu}{2}N$ may depend on $(\Delta/E_F)$ only through the parameter $\mu/E_F$ without additional suppression factors. This should hold for the disc geometry with a hard-wall boundary. Although our result is based on numerical calculations in a finite system, our approach requires very few additional assumptions. 

\subsection{Different confining potentials} \label{potential}

So far we have considered a hard-wall confining potential which vanishes on the disc and goes to infinity outside. In this subsection, we will consider the effects of smooth rotationally symmetric confining potentials. The relation between $L_z$, $N$ and $Q$, embodied by Eqs.~(\ref{what}) and (\ref{Q}), is applicable as long as $\hat{Q}$ commutes with the Hamiltonian. Particularly, $\hat{Q}$ remains a constant of motion for arbitrary rotationally symmetric confining potentials. Therefore one can study the ground state orbital angular momentum through the spectral asymmetry index as done above for an abrupt potential. This is interesting because effects of the shape of the confining potential on the ground state angular momentum or edge currents has been debated in the literature.\cite{stone1, huang, huang2, goryo} 
\begin{figure}
\includegraphics[width=0.8\columnwidth]{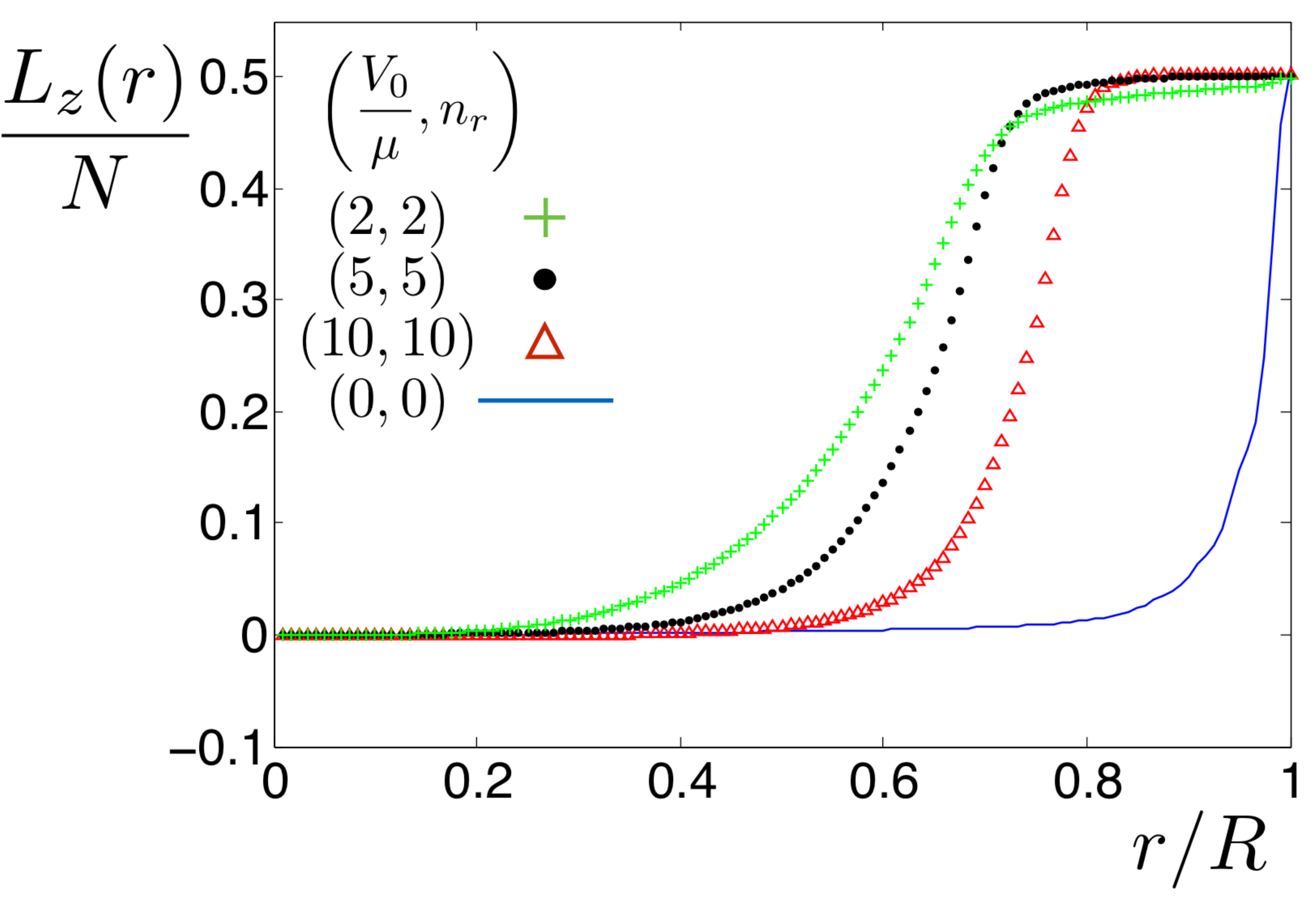}
\caption{Accumulation of orbital angular momentum in the case $\nu=1$ as a function of the radial coordinate. The different curves correspond to different confining potentials labelled by $V_0$ and $n_r$. Even though $N$ and the spatial distribution of $L_z$ vary strongly as the potential is varied, in all the cases we recover the universal result $L_z(R)/N=1/2$. The solid line represents the bare hard-wall boundary condition. All curves are calculated for parameters $R=80k_F^{-1}$ and $k_F\Delta_0/\mu=0.2$. }
\label{rad}
\end{figure}

Here we consider a smooth confinement potential $V(r)=V_0(r/R)^{n_r}$ for $n_r\geq2$ and $V_0 \geq0$ that is imposed on top of the hard-wall potential. In the large $n_r$ limit this potential approaches to an abrupt step. By diagonalising the system for different parameters we find that the spectral asymmetry vanishes for  $\nu=1$, thus the ground state value is universal $L_z=N/2$.  In Fig.~\ref{rad}, we have plotted the accumulation of orbital angular momentum $L_z(r)$ as a function of the radial coordinate. This is calculated by finding the eigenfunctions and evaluating $L_z(r)$ from Eq.~(\ref{lz}) by replacing the upper limit $R$ in the integrals with the radial coordinate $r$.  In the hard-wall case the angular momentum is concentrated within the distance of one coherence length $\xi=v_F/\Delta_0k_F$ from the edge.  The steep potentials $n_r=5,10$ mimic this behaviour, though the effective edge is moved to the point where $V(r)$ crosses the chemical potential. In the quadratic case $n_r=2$, the angular momentum is distributed to a wider annulus. 

\begin{figure}
\includegraphics[width=0.95\columnwidth]{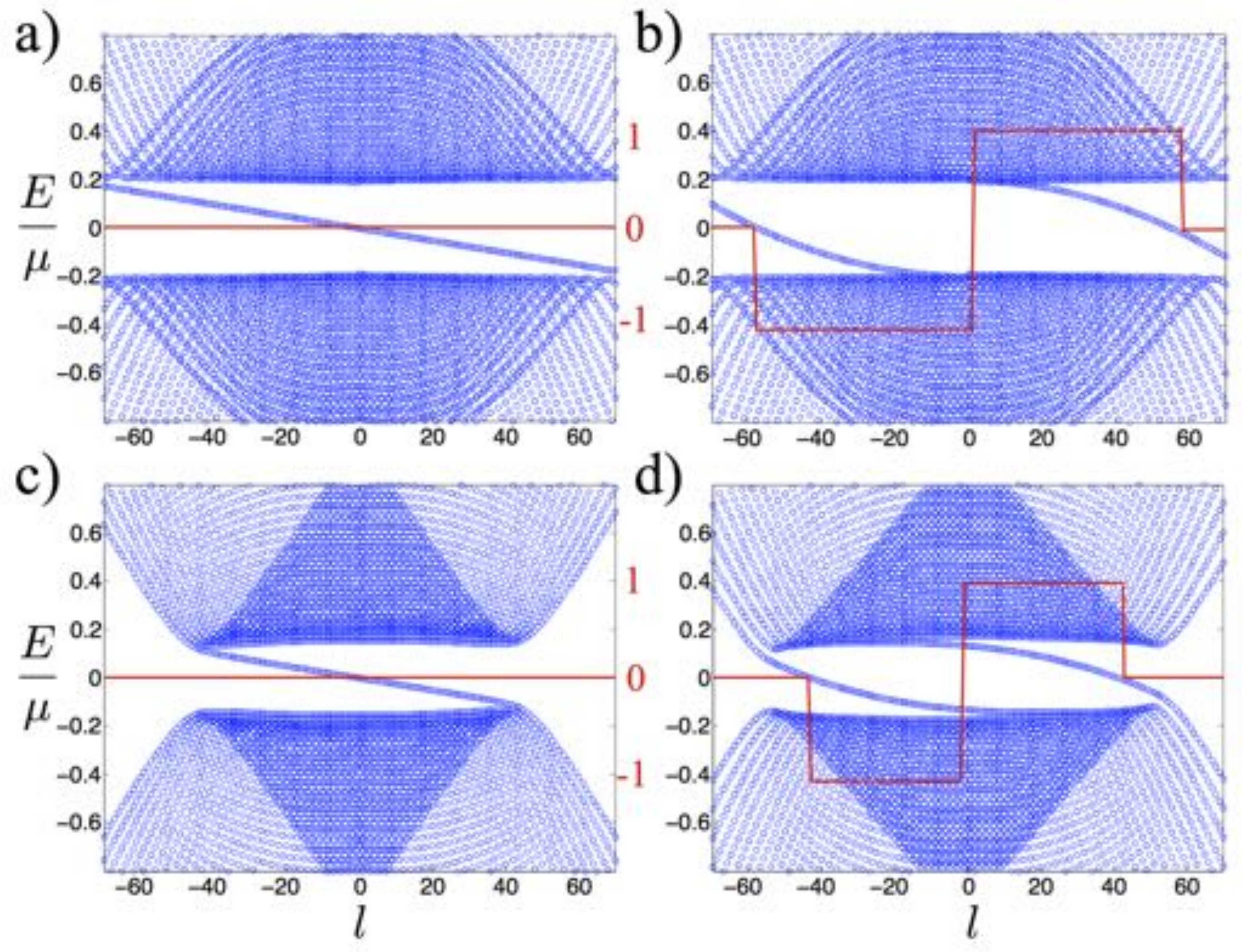}
\caption{Effects of confining potential $V(r)=V_0(r/R)^{2}$ on the spectra of $\nu=1$ (Figs.~a) and c) ) and $\nu=2$ (Figs.~b) and d) ) superfluids. Figures a) and b) depict the spectrum with only the hard-wall potential  $V_0=0$. Figures c) and d) are otherwise the same but with additional quadratic confinement  with $V_0=0.8$. For the $\nu=2$ case the spectral asymmetry changes due to confining potential because the edge modes cross the zero energy at different angular quantum numbers.  All curves are calculated for parameters $R=80k_F^{-1}$ and $k_F\Delta_0^\nu/\mu=0.2$. }
\label{smooth}
\end{figure}
In Fig.~\ref{smooth} we have compared the spectra of the $\nu=1$ and $\nu=2$ pairing states when a smooth confinement potential is added to the hard-wall potential. The spectral asymmetry in the $\nu=1$ case vanishes and the edge branches always cross zero energy at the same point. However, the smooth potential can change the spectral asymmetry in the $\nu=2$ substantially by moving the crossing points of the edge branches. This mechanism is operational in all the $\nu>1$ pairing states and shows that the external potential modifies the ground state orbital angular momentum through the deformation of the edge branches and thus changes the spectral asymmetry of the system. The spectral flow as a function of the shape of the edge potential, leading to the change in the spectral asymmetry for $\nu>1$, was recently reported in a lattice model with Chern number 2 in a cylindrical geometry.\cite{huang2}

\begin{figure}
\includegraphics[width=0.75\columnwidth]{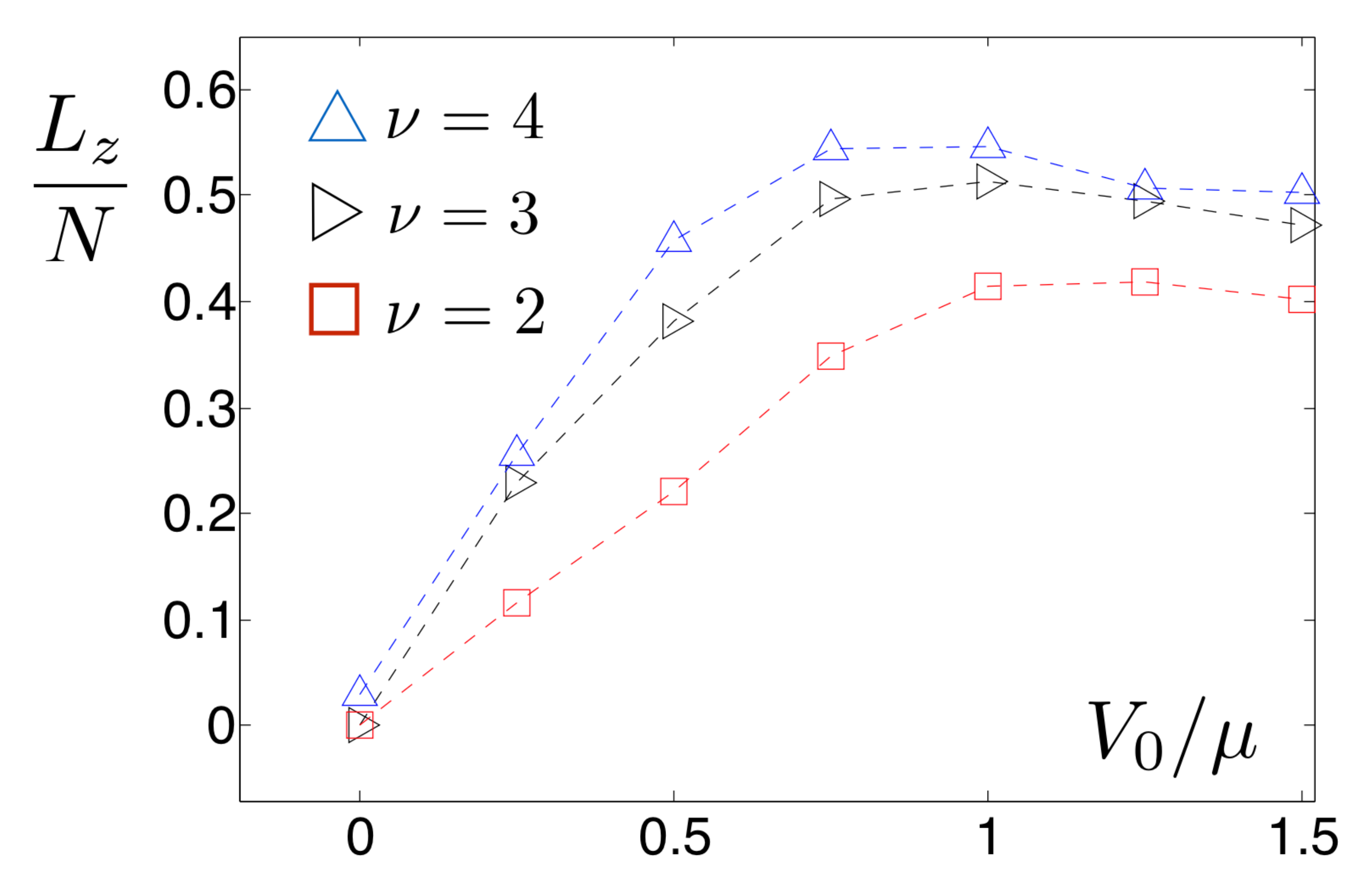}
\caption{Orbital angular momentum per particle in the presence of additional confining potential $V(r)=V_0(r/R)^{4}$. The different curves correspond to different chiral pairing symmetries. At $V_0=0$ only the hard-wall potential is present and $L_z$ is suppressed.  All curves are calculated for parameters $R=80k_F^{-1}$ and $\Delta/\mu=0.1$. }
\label{comp}
\end{figure}
In Fig.~\ref{comp}, we have plotted the angular momentum per particle in the presence of a quartic potential $V(r)=V_0(r/R)^{4}$  for  different pairings $\nu\geq2$. In contrast to the $\nu=1$ case where $L_z=N/2$ remains universal for any $V_0$, $n_r$, the ratio $L_z/N$ varies strongly as a function of $V_0$. As concluded in the previous section, in the case of mere hard-wall potentials ($V_0=0$), angular momentum is strongly suppressed $L_z\approx0$. As $V_0$ is increased, the zero-energy crossings of the edge branches for $\nu\geq 2$ are susceptible to change, therefore also the spectral asymmetry index and  $L_z/N$ vary significantly. This trend is not special for the quartic potential and similar behaviour is recovered for different $n_r$. It is remarkable how large variations in the ratio $L_z/N$ can be achieved by varying the external potential. 

Our results are here are qualitatively in agreement with the ones discussed in Refs.~\onlinecite{stone1, huang} where it was found that the strong suppression of the angular momentum for $\nu>1$ in sharp confining potential is substantially changed when the edge potential is smooth. However, in Ref.~\onlinecite{huang} it was proposed that for potentials that are smooth compared to the coherence length one would obtain the full value $\hat{L}_z=\nu \hat{N}/2$ even for $\nu>1$. Also, similar methods in a strip geometry predicted that edge currents would be universal and determined by the Chern number.\cite{huang2}  As Fig.~\ref{smooth} shows, the spectral asymmetry for $\nu=2$ is diminishing when a smooth potential is added. However, achieving the full value $\hat{L}_z=\nu \hat{N}/2$  would require that the spectral asymmetry vanishes for which, at least in our calculations, does not take place for smooth potentials  and $\nu>1$ studied here.

To conclude this section, we note that the spectral asymmetry index yields a compact answer to the ground state angular momentum in a hard-wall annulus geometry with inner and outer edges at $R_1$ and $R_2$. Provided that the inner radius $R_1$ and the width of the annulus $R_2-R_1$ are larger than the coherence length, both the inner (at $R_1$) and outer (at $R_2$) edges support independent edge states that propagate in opposite directions. The ground state angular momentum is given by $L_z=\nu N/2+Q$ where the spectral asymmetry reduction factor is given by $Q=Q(R_2)-Q(R_1)$,  where $Q(R)=-\nu(k_FR)^2/4$. This result holds for all $\nu>1$ for small gaps $\Delta_0k_F^\nu/E_F\sim 0.1$, and for $\nu=2,3$ with the modification $Q\to Q\frac{\mu}{E_F}$ even for large gaps $\Delta_0k_F^\nu/E_F\sim1$.

\section{Total angular momentum in the presence of spin-orbit coupling and Zeeman field}

\subsection{General considerations}

In this section our aim is to generalize the ideas of spectral asymmetry and the ground state angular momentum of chiral superconductors to the case where the spin and orbital degrees of freedom are coupled. In such systems orbital angular momentum is not a good quantum number even in the absence of pairing, so the spectral asymmetry arguments based on commutation of the operator $\hat{Q}$ and $\hat{H}$ as discussed in the last section are not applicable. Below we will consider a finite Rashba spin-orbit coupling and Zeeman field, as specified by $\vec{b}_{\alpha \beta}$ in Eq.~(\ref{h2}). We have chosen include the Rashba coupling since it has intrinsically two-dimensional nature and is rotationally invariant. 

We will consider triplet pairing $\Delta(\vec{k})=i\sigma_\mu d_\mu(\vec{k})\sigma_y$ specified by
\begin{align} \label{gap1}
\sigma_\mu d_\mu(\vec{k})=&\frac{\Delta_\uparrow}{2}(\sigma_x+i\sigma_y)(k_x+ik_y\,\mathrm{sign}\, \nu_\uparrow)^{|\nu_{\uparrow}|}+\nonumber\\ 
&+\frac{\Delta_\downarrow}{2}(\sigma_x-i\sigma_y)(k_x+ik_y\, \mathrm{sign}\, \nu_\downarrow )^{|\nu_\downarrow|},
\end{align}
which describes a state where up and down spins have different chiral pairing symmetries characterized by odd integers $\nu_{\uparrow}$ and $\nu_{\downarrow}$. We have allowed for the possibility of positive and negative chirality determined by the sign of $\nu_{\uparrow,\downarrow}$. The total angular momentum of a Cooper pair consists of the orbital and spin contributions. The spin up Cooper pairs carry $\nu_{\uparrow}\hbar+2\times\frac{\hbar}{2}=(\nu_{\uparrow}+1)\hbar$ units and the spin down pairs $\nu_{\downarrow}\hbar-2\times\frac{\hbar}{2}=(\nu_{\downarrow}-1)\hbar$ units of angular momentum. For the two spin sectors to transform identically under rotations, we demand that they belong to the same angular momentum channel $j_z=(\nu_{\downarrow}-1)=(\nu_{\uparrow}+1)$. This implies that the winding numbers must satisfy $\nu_{\uparrow}-\nu_{\downarrow}=-2$.

In Sec.~\ref{oam} we employed  the fact that $\hat{Q}$ was a constant of motion so that different eigenstates of $\hat{H}$  could be characterised by its eigenvalues $Q$. This immediately yielded a relation between the ground state expectation value of the orbital angular momentum and the number of particles. In addition, the eigenvalue of $\hat{Q}$ in the ground state could be expressed in terms of the spectral asymmetry of the BdG Hamiltonian. To proceed along those lines we seek to generalize quantity the $\hat{Q}$ that would play the same role for the Hamiltonian (\ref{h1}) with a spin-orbit coupling, Zeeman field and pairing term $(\ref{gap1})$. Therefore we define 
\begin{align} \label{Q2}
\hat{Q}_J=\hat{J_z}-\frac{j_z}{2}\hat{N}
\end{align}
where $\hat{J}_z=\hat{L}_z+\hat{S}_z$ is the total angular momentum operator and $j_z$ is the total angular momentum of Cooper pairs counting spin and orbital contributions.  This operator satisfies 
\begin{align} \label{com}
[\hat{Q}_J,\hat{H}]=0,
\end{align}
where  $\hat{H}$ is given by Eq.~(\ref{h1}) with a finite spin-orbit coupling and the pairing symmetry (\ref{gap1}). This can be seen in the following way. Since $\hat{J}_z$ and $\hat{N}$ commute with $\hat{h}$ in Eq.~(\ref{h2}), it follows that $[\hat{Q}_J,\hat{H}]=[\hat{Q}_J,\hat{\Delta}]$.  The  structure (\ref{gap1}) implies that the pairing operator has two spin contributions $\hat{\Delta}=\hat{\Delta}_{\uparrow\uparrow}+\hat{\Delta}_{\downarrow\downarrow}$, which in momentum representation take the form $\hat{\Delta}_{\uparrow\uparrow}=\hat{\Omega}_{\uparrow}+\hat{\Omega}^\dagger_{\uparrow}$,  $\hat{\Delta}_{\downarrow\downarrow}=\hat{\Omega}_{\downarrow}+\hat{\Omega}^\dagger_{\downarrow}$ with  $\hat{\Omega}_{\sigma}=\sum_kf_\sigma(k)\hat{a}_{k\sigma}\hat{a}_{k\sigma}$. Here $f_\sigma(k)=\Delta_\sigma(k_x+ik_y\,\mathrm{sign}\,\nu_\sigma)^{|\nu_\sigma|}$ contains the orbital structure of the pairing of different spin projections.  Now one can prove that the individual commutators  satisfy $[\hat{L}_z,\hat{\Omega}_{\sigma}]=\nu_\sigma\hat{\Omega}_{\sigma}$, $[\hat{S}_z,\hat{\Omega}_{\sigma}]=(\delta_{\sigma \uparrow}-\delta_{\sigma \downarrow})\hat{\Omega}_{\sigma}$ and $[\hat{N},\hat{\Omega}_{\sigma}]=2\hat{\Omega}_{\sigma}$. Collecting these results and recalling that $j_z=(\nu_{\downarrow}-1)=(\nu_{\uparrow}+1)$, we arrive at the conclusion $[\hat{Q}_J,\hat{H}]=0$.

The above result establishes that the separate symmetry $U(1)_{J_z}\times U(1)_{N}$ of a rotationally invariant and particle conserving system reduces to a single combined symmetry $U(1)_{J_z-j_zN/2}$ due to the pairing term. All the eigenstates of (\ref{h1}) are labelled by eigenvalues of $Q_J$ of the operator $\hat{Q}_J$ and the ground state expectation value of total angular momentum satisfies  $J_z=\frac{j_zN}{2}+Q_J$. Analogously to the case in the previous section, $Q_J$ can be regarded as the reduction of the total angular momentum from the value gained by summing up angular momenta of the Cooper pairs. This reduction is, again, associated with the edge states circulating around the disc. The edge states are guaranteed to exist at least when the Chern number is nonzero. In the absence of the spin-orbit coupling, the Chern number of the studied model is given by $C=\nu_{\uparrow}\theta(\mu-B_z)+\nu_{\downarrow}\theta(\mu+B_z)$, where $\theta(x)$ is the Heaviside step function. This result also holds in the presence of the Rashba coupling as long as the energy gap is not closed. Edge states may also exist for vanishing Chern number, for example
the case $\nu_{\uparrow}=-1$, $\nu_{\downarrow}=1$ with $B_z=0$ and $\Delta_\uparrow=\Delta_\downarrow$ corresponds to a time-reversal invariant topological superconductor with a $\mathbb{Z}_2$ invariant.\cite{qi2}  

As discussed in Sec.~\ref{oam}, the ground state of the superfluid (the positive-energy vacuum) can be constructed by filling the negative-energy Bogoliubov quasiparticle states obtained as a solution of Eq.~(\ref{bdg2}). The unitary transformation $U(\phi)$ that makes Eq.~(\ref{bdg2}) separable is now fixed by $N_1=j_z-1$, $N_2=j_z$, $N_3=0$, $N_4=-1$. One can show that each negative-energy state $\psi_{nl}$ gives a contribution of $(l+\frac{j_z}{2}-\frac{1}{2})/2$ to the ground-state eigenvalue $\hat{Q}_J$, so we can write  $Q_J=\sum_{E_{n,l}<0}(l+\frac{j_z}{2}-\frac{1}{2})/2$. Because the BdG Hamiltonian (\ref{bdg2}) in this case satisfies particle-hole symmetry $P H(r,l)P^{-1}=-H^*(r,-l-j_z+1)$, where $P=i\tau_y$, the negative and positive eigenvalues are related by $E_{n,l}=-E_{n,-l-j_z+1}$. With the help of this relation we can write the ground state eigenvalue  $\hat{Q}_J$ as 
\begin{align} \label{Q3}
Q_J=-\frac{1}{2}\sum_{l} (l+\frac{j_z}{2})\eta_l,
\end{align}
where   $\eta_l=\frac{1}{2}\sum_n \text{sign}\,E_{n,l}$ is the spectral asymmetry index  expressed in terms of eigenvalues  of the  BdG Hamiltonian (\ref{bdg2}). The relations (\ref{Q2}), (\ref{com}) and (\ref{Q3}) are the main results of this section and together provide a framework for studying the total angular momentum in the presence of spin-orbit coupling in triplet superfluids with the pairing symmetry (\ref{gap1}). Below we will study two examples that highlight the distinct behaviour of the cases $j_z=0$ and $j_z\neq0$.

\subsection{ Ground state angular momentum with the $\nu_{\uparrow}=-1$, $\nu_{\downarrow}=1$ pairing }

\begin{figure}
\includegraphics[width=0.99\columnwidth]{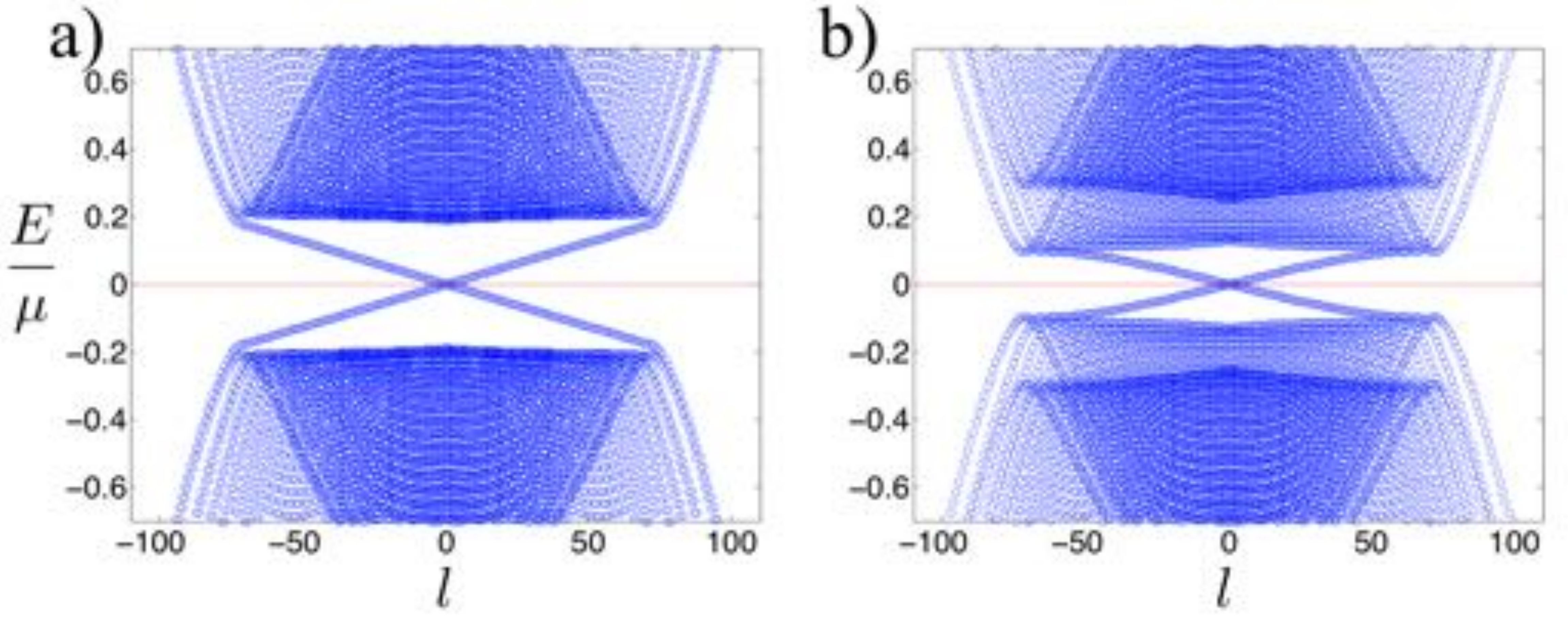}
\caption{Spectrum of the $\nu_{\uparrow}=-1$, $\nu_{\downarrow}=1$ superfluid as a function of the angular quantum number $l$ without Rashba coupling (a) and with a finite Rashba coupling $\alpha k_F/\mu=0.1$ (b).   Both figures are calculated for parameters $R=80k_F^{-1}$ and $k_F\Delta_0/\mu=0.2$ and $B_z=0$. The spectral asymmetry (red line) vanishes in both cases. }
\label{nu1nu_1}
\end{figure}

\begin{figure}
\includegraphics[width=0.7\columnwidth]{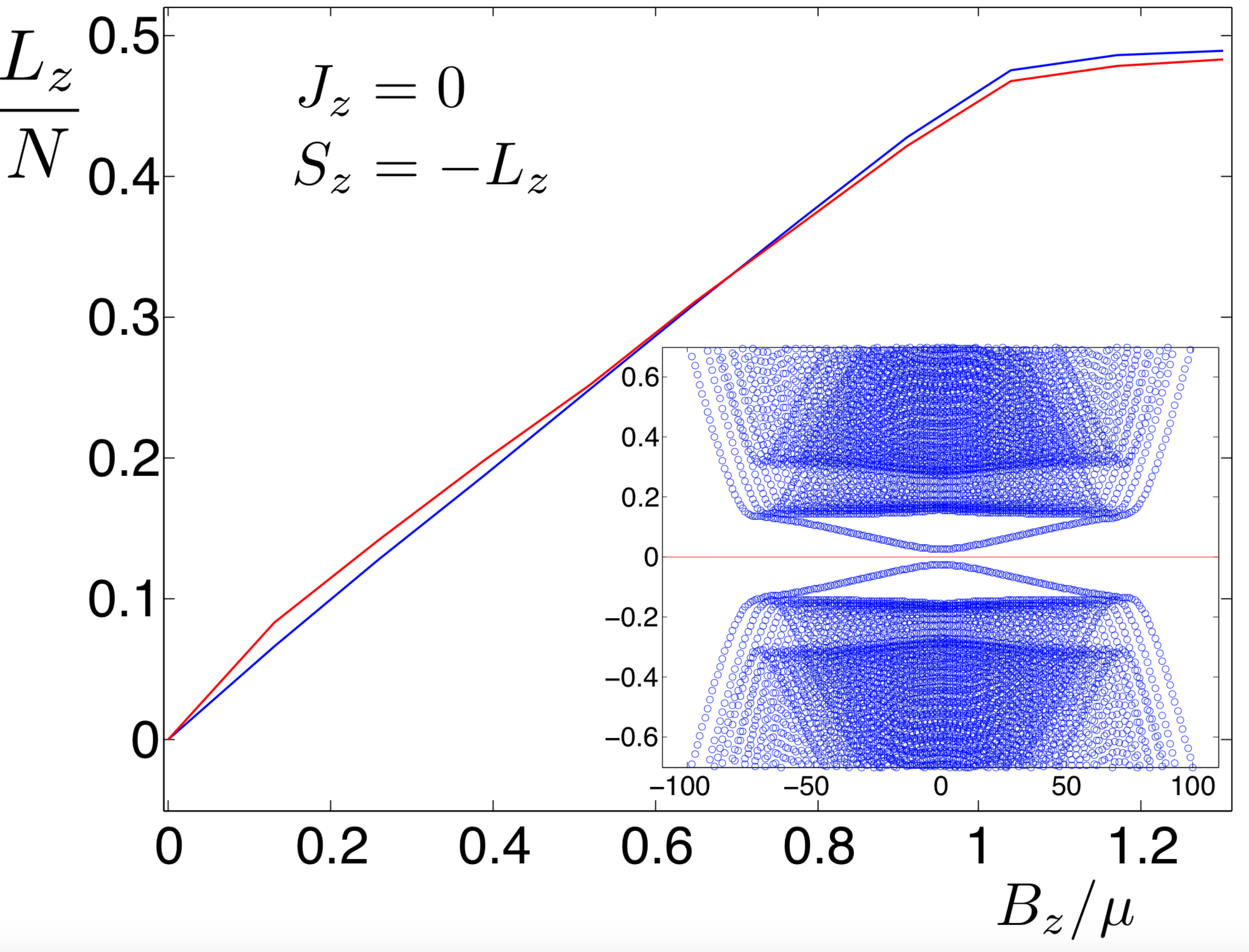}
\caption{Angular momentum of the $\nu_{\uparrow}=-1$, $\nu_{\downarrow}=1$ case as a function of magnetic field. Different curves correspond to different values for the Rashba constant $\alpha k_F=0$ (blue), $\alpha k_F=0.3\mu$ (red). For both curves $\Delta_0 k_F=0.3\mu$. Inset: same as Fig.~\ref{nu1nu_1} b) but with a finite Zeeman field $B_z=0.1\mu$. }\label{lz1}
\end{figure}

We now turn to study the model (\ref{h1}) with pairing symmetry (\ref{gap1}) specified by winding numbers $\nu_{\uparrow}=-1$ and $\nu_{\downarrow}=1$. In a finite magnetic field $B_z$ but with vanishing Rashba coupling $\alpha=0$ the system consists of two noninteracting chiral $p_x\pm ip_y$ superconductors  with counter-propagating edge states and spin-dependent chemical potentials $\mu\pm B_z$. In the presence of spin-orbit coupling the counter propagating edge states persist as long as  $B_z=0$ and, even though the Chern number is zero. The spectrum of the system is illustrated in Fig.~\ref{nu1nu_1}. Only the combined effect of the Zeeman field and Rashba coupling will lift the crossing of the edge modes, as illustrated in the inset of Fig.~\ref{lz1}.

Since the total angular momentum of the Cooper pairs vanishes $j_z=0$, Eq.~(\ref{Q2}) states that the total angular momentum is given by the spectral asymmetry $J_z=Q_J$. Because the system exhibits  edge states, it is possible that these may give rise to a finite spectral asymmetry. However, in analogy to the case $k_x+ik_y$ studied in the previous section, the numerical evaluation of the spectral asymmetry (\ref{Q3}) reveals that $Q_J=0$ even for a finite spin-orbit coupling and Zeeman field. Therefore we conclude that  $J_z=0$. It should be noted that the ground state is also an eigenstate of  $\hat{J}_z$. The vanishing $J_z$ in turn implies that the ground state expectation values of orbital and spin angular momentum cancel $L_z=-S_z$. By applying a Zeeman field, the orbital angular momentum also  changes to cancel the spin polarization. 

By calculating the orbital angular momentum and spin from Eq.~(\ref{lz}) we have directly verified the prediction $L_z=-S_z$ of the spectral asymmetry argument. In Fig.~\ref{lz1} we have plotted the expectation value of the ground state orbital angular momentum as a function of the Zeeman field $B_z$.  When $B_z=0$ and  $\Delta_\uparrow=\Delta_\downarrow=\Delta_0$, the of number particles with up and down coincide and $L_z=0$. Since $B_z$ acts as a spin-dependent chemical potential, by increasing $B_z$  we increase particles with $k_x+ik_y$ pairing and decrease the $k_x-ik_y$ component, hence driving the system into the $L_z/N>0$ regime.  Since the $k_x-ik_y$ component is effectively suppressed when $B_z>\mu$ we expect that the system approaches the value $L_z/N= \frac{1}{2}$ for large $B_z$, in accordance with the findings of the previous section. As Fig.~\ref{lz1} illustrates, that is indeed the case. At strong spin-orbit coupling the limiting value is approached slower than in the absence of it. At $B_z=\mu$ the effective chemical potential of spin up particles is zero and the the system undergoes a topological phase transition between phases with Chern numbers 0 and 1. The transition shows up in the figure as a crossover between two trends at $B_z=\mu$.

\subsection{ Ground state angular momentum with the $\nu_{\uparrow}=1$, $\nu_{\downarrow}=3$ pairing }

\begin{figure}
\includegraphics[width=0.99\columnwidth]{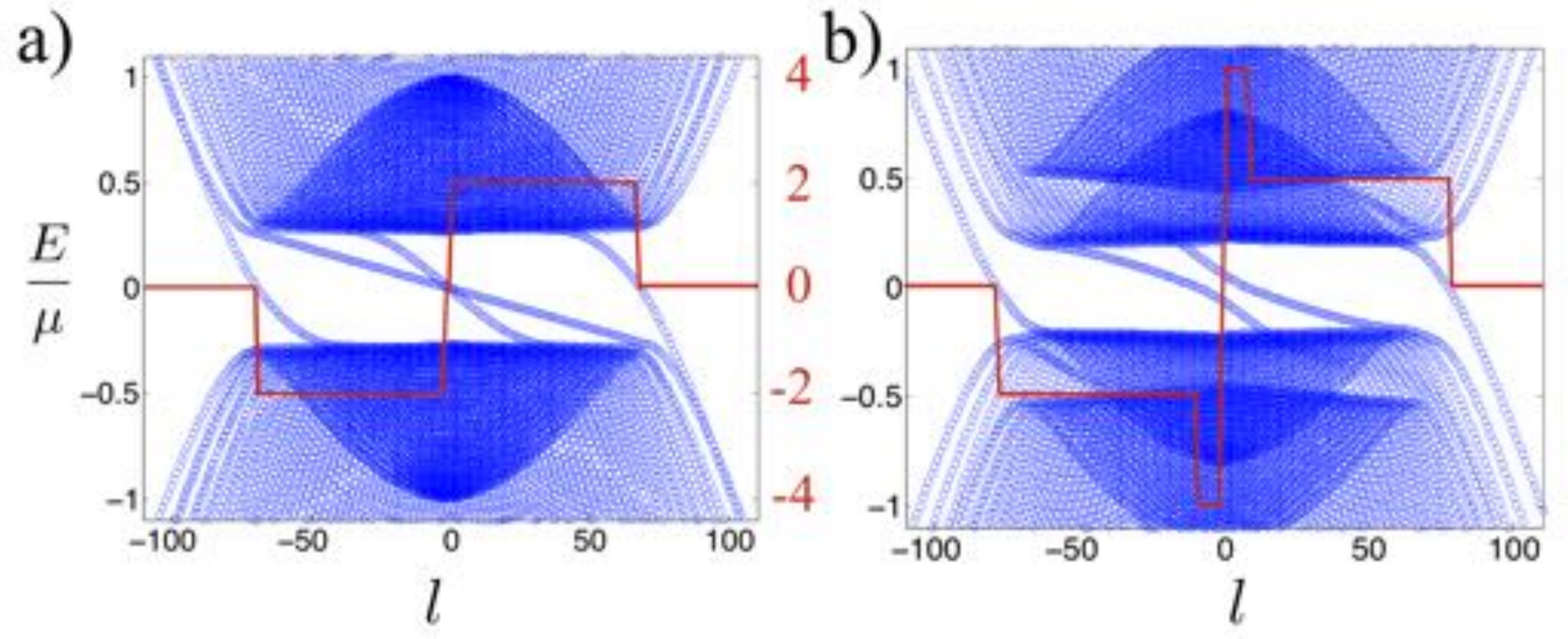}
\caption{a): Spectrum of the $\nu_{\uparrow}=1$, $\nu_{\downarrow}=3$ superfluid as a function of the angular quantum number $l$ for $\alpha k_F=0$ and $B_z=0$.  b): The same as a) but for  $\alpha k_F=0.2\mu$, $B_z=0.2\mu$ . Both figures correspond to parameters  $R=80k_F^{-1}$, $\Delta_\uparrow/\mu=0.2$ and $\Delta_\downarrow/\mu=0.3$. The spectral asymmetry is represented by the red line with the associated red scale. }
\label{nu1nu3}
\end{figure}

Now we will study the properties of  system (\ref{h1}), with pairing symmetry (\ref{gap1}) and winding numbers $\nu_{\uparrow}=1$, $\nu_{\downarrow}=3$. In this case Cooper pairs carry total angular momentum $j_z=2$. This type of $j_z=2$ pairing is proposed to take place in a parallel-plane geometry in $^3$He-A$_1$.\cite{volovik6}  The spectrum of the  system is illustrated in Fig.~\ref{nu1nu3}. For weak Zeeman field, the Chern number is four and the gap is traversed by four edge states with the same chirality. While the Rashba coupling modifies the spectrum quite drastically it cannot destroy the edge states without closing the bulk energy gap. 

In contrast to the $\nu_{\uparrow}=-1$, $\nu_{\downarrow}=1$ case above, Fig.~\ref{nu1nu3} shows that the spectral asymmetry and thus $Q_J$ is finite. Therefore the relation between $J_z$ and $N$ is nontrivial, in analogy to the relation between $L_z$ and $N$ in the case of chiral pairing (\ref{delta1}) for $\nu>1$. By solving the BdG eigenvalue problem (\ref{bdg2}), we evaluate the angular momentum reduction factor in the ground state by employing Eq.~(\ref{Q3}). The relation between $J_z$ and $N$ then follow from Eq.~(\ref{Q2}). 

\begin{figure}
\includegraphics[width=0.99\columnwidth]{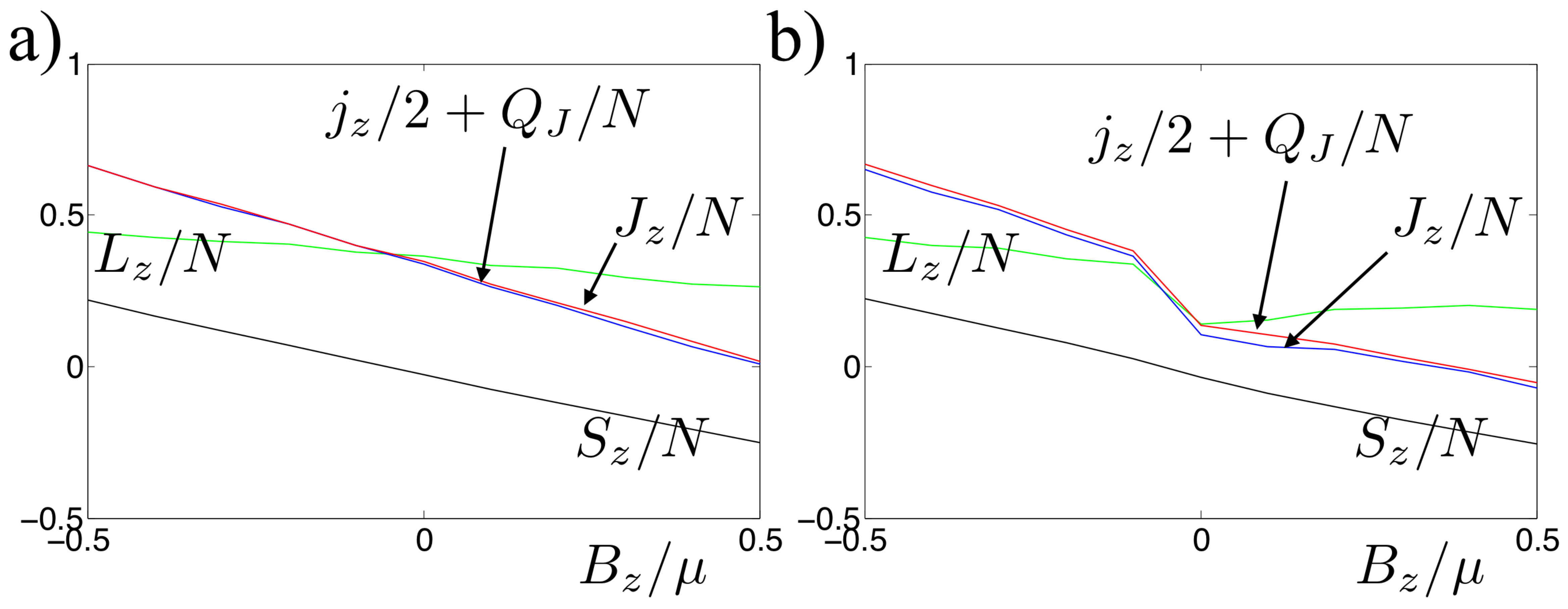}
\caption{ a): Different angular momentum contributions and the reduction factor $Q_J/N$ in the case $\nu_\uparrow=1, \nu_\downarrow=3$  ($j_z=2$) as a function of the Zeeman field for  a vanishing spin-orbit coupling.  b): The same as a) but for a finite spin-orbit coupling $\alpha k_F=0.25\mu$. In both figures $k_F\Delta_\uparrow=0.3\mu$, $k_F^3\Delta_\downarrow=0.3\mu$ and $R=80k_F^{-1}$. }
\label{J2}
\end{figure}

In Fig.~\ref{J2} we illustrate how $L_z/N$, $S_z/N$ and $J_z/N$, calculated by employing Eq.~(\ref{lz}), depend on the Zeeman field. We have also plotted the quantity $\frac{j_z}{2}+Q_J/N$ which is determined by the spectral asymmetry and should coincide with $J_z/N$ according to Eq.~(\ref{Q2}). The two methods for calculating $J_z$ agree within the accuracy of our numerics. As Figs.~\ref{nu1nu3} and \ref{J2} show, the spectral asymmetry and the total angular momentum in the ground state also varies significantly as a function of the Zeeman field, in contrast to the $\nu_{\uparrow}=-1$, $\nu_{\downarrow}=1$  case  where they always vanish. By comparing the total angular momentum with and without the spin-orbit coupling in Fig.~\ref{J2} a) and b) one can conclude that the expectation value of the spin seems to be unaffected by the spin-orbit coupling. 

The two different superfluids $\nu_{\uparrow}=-1$, $\nu_{\downarrow}=1$ and $\nu_{\uparrow}=1$, $\nu_{\downarrow}=3$ illustrate that, analogously to the different behaviour of orbital angular momentum for $\nu=1$ and $\nu>1$, the total angular momentum also behaves qualitatively differently depending on whether $|\nu_\sigma|=1$ or $|\nu_\sigma|>1$.

\section{ Discussion}

The ground state angular momentum in chiral superfluids has been studied for 40 years and the recent advances in topological superconductivity and the connections to Hall viscosity\cite{read2, shitade2} have revived the interest in the subject. Recent developments have also brought new aspects into focus. As highlighted in Ref.~\onlinecite{tada} for different chiral pairing symmetries, angular momentum in a finite geometry is not a topological property of the ground state. Nevertheless, in the models studied in our work, topology necessitates the existence of the edge states that give rise to a finite spectral asymmetry. The connection between angular momentum and the spectral asymmetry arises from the fact that the studied Hamiltonians allow for a conserved quantity $\hat{Q}=\hat{L}_z-\nu \hat{N}/2$ or $\hat{Q}_J=\hat{J}_z-j_z\hat{N}/2$. In this case the ground-state expectation value of $\hat{Q}$ (or $\hat{Q}_J$) may change only through a spectral flow. The spectral flow is caused by a topological phase transition or a deformation of the  edge branches in a way that an eigenstate of the Hamiltonian crosses zero energy. The connection between angular momentum and spectral flow is, however, absent for example in the artificial $p$-wave model studied in Ref.~\onlinecite{shitade1} by Shitade and Nagai.  They found that in the artificial $p$-wave model, achieved through the interplay of the $s$-wave proximity effect, the Rashba effect and Zeeman field,  angular momentum changes continuously even when the gap remains open and no states cross the zero energy. Orbital angular momentum exhibits a smooth crossover behaviour at the topological phase transition between a trivial and a finite Chern number states.  The qualitative difference  between the model studied in Ref.~\onlinecite{shitade1} and the models studied in this work arises from the existence of the conserved quantity $\hat{Q}$ that strongly constrains the behaviour of the system.   

The special behaviour of angular momentum for $(p_x\pm ip_y)^\nu$ pairing with $\nu=1$ compared to $\nu>1$ case with and without spin-orbit coupling is remarkable. As discussed above, the universality of the result $L_z=\pm\hbar N/2$ irrespective of different confining potentials and pairing strength can be understood in terms of vanishing spectral asymmetry.  The vanishing spectral asymmetry results from the fact that the edge spectrum is monotonic and passes through the gap center at zero angular momentum.  The topology and the particle-hole symmetry of the BdG spectrum do not prohibit a deformation of the edge spectrum in a way that would result in a finite spectral asymmetry.\cite{volovik2} Also, the different mechanisms of suppression of angular momentum has been proposed in $\nu=1$ in Refs.~\onlinecite{mizushima1, mizushima2}.  However, as discussed in Sec.~\ref{potential}, deformations of the radial confining potential do not lead to finite spectral asymmetry or departures of the angular momentum from the universal value.

\section{ Summary }

In the present work we studied the ground state angular momentum in chiral superfluids and superconductors confined to a disc. Motivated by the recent discovery that orbital angular momentum in chiral $(p_x+ip_y)^\nu$ systems behaves dramatically differently for $\nu=1$ and $\nu>1$, we examined several further directions. By employing spectral asymmetry arguments we showed that for $\nu=1$, the result $L_z=\hbar N/2$ is robust against deformations of the confining potential, appropriate for smooth and hard-wall potentials as well as an annulus geometry. In contrast, for $\nu>1$ the ratio $L_z/N$ can be strongly tuned by an external rotationally symmetric potential.  Also, our results imply that the angular momentum reduction from the naive value $\nu N/2$ depends on the gap size only through parameter $\frac{\mu}{E_F}$  for $\nu=2,3$ but for $\nu\geq4$ behaves qualitatively differently. thus, for $\nu=2,3$ pairing states in a hard-wall geometry, the ground state angular momentum is suppressed as $L_z=-\frac{\nu}{2}\hbar N(1-\frac{\mu}{E_F}) $ and no additional suppression factors $(\frac{\Delta}{E_F})^\gamma$ arise.

In the second part of the work we studied chiral superfluids where $L_z$ is not a good quantum number in the normal state due to a Rashba-type spin-orbit coupling. Inspired by previous ideas by Volovik\cite{volovik3} and Tada \emph{et al.},\cite{tada} we formulated a new relation $J_z=j_zN/2+Q_J$ between the total angular momentum $J_z$, the number of particles $N$ and the reduction factor $Q_J$ determined by the spectral asymmetry index of the Bogoliubov-de Gennes Hamiltonian on a disc. This relation holds for the triplet states where Cooper pairs have a different odd pairing symmetry but the same total angular momentum $j_z$. We discovered that, in analogy to the orbital angular momentum dichotomy between the $\nu=1$ and $\nu>1$ cases for the $(p_x+ip_y)^\nu$ pairing, the behaviour of the spectral asymmetry and total angular momentum is different depending on whether $j_z=0$ or $j_z\neq 0$. 

\acknowledgments
The author would like to thank G. E. Volovik for directing his interest in the studied problem and many valuable discussions along the way. The author acknowledges Alex Weststr\"om and Kim P\"oyh\"onen for discussions and valuable comments on the manuscript. This work was supported by the Academy of Finland.

\end{document}